\documentclass[12pt,preprint]{aastex}

\shorttitle{3D LISM Morphology}
\shortauthors{}

\begin{document}

\newcommand{\php}[0]{\phantom{--}}
\newcommand{\kms}[0]{km~s$^{-1}$}

\title{Evaluating the Morphology of the Local Interstellar Medium: 
Using New Data to Distinguish Between Multiple Discrete Clouds and a Continuous Medium
\footnote{Based on observations made with the NASA/ESA
Hubble Space Telescope, obtained from the Data Archive at the Space
Telescope Science Institute, which is operated by the Association of
Universities for Research in Astronomy, Inc., under NASA contract NAS
AR-09525.01A. These observations are associated with programs \#11568.}}

\author{Seth Redfield\altaffilmark{2} and Jeffrey L. Linsky\altaffilmark{3}}

\altaffiltext{2}{Astronomy Department and Van Vleck Observatory,
Wesleyan University, Middletown, CT 06459-0123; {\tt sredfield@wesleyan.edu}}

\altaffiltext{3}{JILA, University of Colorado and NIST, 
Boulder, CO 80309-0440; {\tt jlinsky@jila.colorado.edu}}

\begin{abstract}
Ultraviolet and optical spectra of interstellar gas along the
lines of sight to nearby stars have been interpreted 
 by \citet{Redfield2008} and previous studies as a set of
discrete warm, partially ionized clouds each with a different flow
vector, temperature, and metal depletion. Recently, \citet{Gry2014}
have proposed a fundamentally different model consisting of a 
single cloud with nonrigid flows
filling space out to 9 parsecs from the Sun that they 
propose better describes the local ISM. Here we test these 
fundamentally different morphological 
models against the spatially unbiased
\citet{Malamut2014} spectroscopic data
set, and find that the multiple cloud morphology model provides a better fit 
to both the new and old data sets. The detection of three or more
  velocity components along the lines of sight to many nearby stars,
the presence of nearby scattering screens, the observed thin elongated 
structures of warm interstellar gas, and the likely presence of strong
interstellar magnetic fields also 
support the multiple cloud model. The
detection and identification of intercloud gas and the measurement of
neutral hydrogen density in clouds beyond the Local Interstellar
Cloud could provide future morphological tests. 

\end{abstract}


\section{DEVELOPMENT OF MULTIPLE CLOUD MORPHOLOGIES}

The interstellar medium plays a critical role in astrophysics as the 
interface between the creation of metals in stars and their eventual
inclusion in the next generation of stars. A significant fraction of the mass 
of galaxies is 
interstellar gas and the accretion and loss of this gas play  
critical roles in galactic evolution \citep[e.g.,][]{Kennicutt1998}. 
Studies of the physical processes,
thermal structure, and dynamics of interstellar gas include 
theoretical models, numerical simulations, and 
extensive observational studies typically involving
high-resolution spectra of ultraviolet absorption lines.  The local
interstellar medium (LISM) which exists in close proximity to the Sun
(i.e., within 100 pc), is the closest and simplest sample 
of interstellar gas to study
with high-sensitivity absorption line spectra.  A significant
advantage of using LISM observations to evaluate interstellar structures 
is the small number of traversed absorbers and high likelihood that 
 the interstellar gas velocity structure
along the line of sight will be resolved in a 
 high-resolution stellar spectrum.  

From an analysis of high-resolution \ion{Ti}{2} $\lambda$3384 absorption line 
spectra toward stars within 100~pc of the Sun, \citet{Crutcher1982} found 
that warm gas in the LISM is moving coherently with a heliocentric 
velocity vector
($l_{\circ},b_{\circ},V_{\circ})=(25^{\circ},+10^{\circ},-28$ km s$^{-1}$).
He noted that this flow is consistent with an expanding shell of gas 
accelerated by the hot stars and supernovae in the Sco-Cen Association, and 
that this interstellar wind flow could explain the velocity of resonantly 
scattered solar H~I Lyman-$\alpha$ and He~I resonance lines.

Further understanding of the morphology and kinematics of the LISM required
more sensitive high-resolution spectra, in particular using ultraviolet
absorption lines of \ion{H}{1} and \ion{D}{1} Lyman-$\alpha$, \ion{Mg}{2} $\lambda$2796,2802, 
\ion{Fe}{2} $\lambda$2600, \ion{O}{1} $\lambda$1302,1304,1306, \ion{C}{2} $\lambda$1334,1335,
and many weaker transitions of neutral and singly-ionized elements. 
These spectral lines are particularly useful because the ions 
are abundant in the ISM and the permitted 
transitions are from the highly populated 
ground states. Such data began to appear in quantity from the 
increasingly 
sensitive and higher resolution spectrographs on {\em Copernicus},
{\em International Ultraviolet Explorer} ({\em IUE}), the Goddard High Resolution Spectrograph (GHRS) instrument on the {\em Hubble Space Telescope} ({\em HST}), and finally the 
Space Telescope Imaging Spectrograph (STIS) instrument on {\em HST}. 

Using {\em Copernicus}, \citet{McClintock1978}
was able to measure \ion{H}{1} column densities and radial velocities of 
interstellar gas in the lines of sight toward eight 
G and K stars within 15~pc of the Sun supporting the model of 
uniform flow of neutral interstellar gas in the solar neighborhood. 
{\em IUE} spectra of the \ion{H}{1} Lyman-$\alpha$ line toward many hot stars
allowed \citet{Frisch1983} to create maps of hydrogen column density,
which show an asymmetrical structure with an $N($\ion{H}{1}$)$ hole 
centered at $l\sim 225^{\circ}, b\sim -15^{\circ}$. 

Analysis of high-resolution GHRS spectra of the nearby 
stars $\alpha$~Cen \citep{Linsky1996}, Sirius \citep{Lallement1994,Bertin1995},
Procyon \citep{Linsky1995}, Capella \citep{Linsky1993}, and others provided
accurate column densities, velocities and nonthermal broadening parameters 
for these 
lines of sight by combining information from the saturated \ion{H}{1} Lyman-$\alpha$
line, the thermally-broadened but unsaturated \ion{D}{1} Lyman-$\alpha$ line, and
lines of heavy elements like \ion{Mg}{2} and \ion{Fe}{2} that reveal the velocity structure
along the line of sight.

The next step in our understanding of the structure of partially-ionized 
gas in the LISM
was the recognition that individual comoving structures (called clouds) 
could be identified by the common space velocities of gas 
across large regions of the sky. 
Early efforts used \ion{Ca}{2} observations \citep[e.g.,][]{Lallement1986}, although these cloud vectors were significantly revised with a larger sample of observations \citep{Vallerga1993}.
\citet{Lallement1992}, using essentially \ion{Ca}{2} observations, and \citet{Lallement1995}, using primarily high-resolution ultraviolet spectra, identified 
two clouds: the G cloud in the direction of the Galactic Center, 
and the Local Interstellar Cloud (LIC) centered in the opposite direction 
and likely containing the Sun on the basis of their projected
velocity differences. \citet{Redfield2000} produced a 
three-dimensional map of the LIC on the basis 16 lines of sight observed 
by GHRS, 13 lines of sight observed with the Ca~II lines, 
and three lines 
of sight to hot white dwarfs observed by the {\em Extreme Ultraviolet Explorer} ({\em EUVE}). The accumulation of 
additional observations with the GHRS and STIS instruments on
{\em HST} by \citet{Dring1997,Frisch2002,Redfield2002,Redfield2004} and others
were the basis for more detailed models of the LISM. \citet{Frisch2002} 
proposed a kinematic model with seven clouds based on 96 velocity components
detected toward 60 stars.

\citet{Redfield2008} [hereafter RL08] next developed a 15 cloud model 
of the LISM on the basis of 270 radial velocity
measurements toward 157 stars located within 100~pc of the Sun. 
The assignment of individual velocity components to a cloud was
based mainly on kinematics (i.e., the radial velocities are consistent 
to within the measurement errors of a common velocity vector for a
number of targets distributed over a large region of the sky),
although contiguity of the sight line coordinates was also a
criterion. The shape of the clouds, which was drawn by eye to include the
Galactic coordinates of the assigned targets, is subjective and has
been refined over time (e.g., \citealt{Malamut2014} 
[hereafter M14]) with the inclusion of new sight lines. 
Because of the close proximity of these clouds to the Sun, 
they typically subtend 
large angles on the sky.  This provides leverage to measure the 
three-dimensional velocity vector from a sample of radial velocity 
measurements, and thereby to measure the transverse velocities of these 
clouds.   Clouds at a greater distance, or with small linear sizes
cannot be identified
with this technique. The conditions for which a cloud can be
identified are complex and hard to quantify, but the accuracy of the
velocity vector can be estimated from the errors in its components
(see Table 16 of RL08).

All of these warm partially-ionized clouds, 
now called the complex of local 
interstellar clouds (CLIC), lie entirely or in part within 15~pc 
of the Sun because the nearest 
stars showing absorption by the gas in each cloud lie within this 
distance. Since all of these sight lines show absorption by 
partially-ionized gas in these clouds or in not yet identified clouds, there 
is presently no direction in space where $\log N($\ion{H}{1}$)$ $<17.4$ 
toward an observed star. Comparison of the 
widths of absorption lines of a low mass element (e.g., D) 
with high-mass elements 
(e.g., Fe and Mg) allowed \citet{Redfield2008} to infer the gas temperature
of the LIC to be $T=7500\pm 1300$~K and the temperatures of the other clouds 
to lie in the range 3900~K (Blue) to 9900~K (Mic). 
\citet{Frisch2009} and \citet{Frisch2011} noted that 
the inferred temperatures and nonthermal broadening parameters assume
a Maxwell-Boltzman distribution of velocities and mass-independent turbulence,
both of which may not be valid in low density clouds.

\citet{Frisch2009} concluded that the physical properties of 
the CLIC clouds are
typical of warm partially-ionized gas observed elsewhere in the solar 
neighborhood on the basis of temperature, velocity, composition, ionization, 
and magnetic field properties. In particular, the ionization equilibrium
of the CLIC gas is consistent with the local EUV radiation field
\citep{Slavin2008}.
The recent comprehensive review of the interstellar medium surrounding the 
Sun by \citet{Frisch2011} describes our present understanding of
the Galactic environment of the LISM,
the role of outflowing gas from the Sco-Cen association as a driver for the
CLIC kinematics, 
the interstellar radiation field, ionization and depletion of metals in the 
gas, kinematics of the gas, and the interstellar magnetic field.

\section{AN ALTERNATIVE MODEL FOR THE CLIC MORPHOLOGY}

\citet{Gry2014} [hereafter GJ14] have proposed a fundamentally
different morphological model
for the warm interstellar gas located near the Sun in which all of the space 
within about 9 parsecs of the Sun consists of a single continuous cloud 
with a nonrigid  flow and a gradient in metal depletion properties. 
They fit the \citet{Redfield2002} \ion{Mg}{2} and \ion{Fe}{2} 
radial velocity data set with a nonrigid flow that is differentially
decelerated in the direction of motion and expanding in directions
perpendicular to the flow. In their Single Local Cloud model, 
the flow speed and direction near the Sun is consistent with the speed and
direction of interstellar helium flowing into the heliosphere as 
measured by the {\em Interstellar Boundary Explorer} ({\em IBEX}) and {\em Ulysses} satellites. They found that
their single local cloud model fits nearly all of the \ion{Mg}{2} and \ion{Fe}{2} 
velocity components 
that RL08 assigned  to the LIC, G, NGP, Blue, Leo, Aur, and Cet clouds. 
If the mean neutral hydrogen density is 0.055 cm$^{-3}$, 
as inferred from the hydrogen column densities and
distances of nearby stars, then their single cloud model
fills all of the space out to
roughly 9 parsecs from the Sun. GJ14 also found that another set of 
velocity components that RL08 had assigned to the Hyades and Mic clouds 
and including  many of the previously unassigned velocity components, could
be fit by a second vector that they called the Cetus Ripple, which
may be a signature of a shock front inside of the local cloud. 
They speculated that the remaining
unassigned velocity components are also perturbations located inside 
of the local cloud.

\section{A CRITICAL TEST OF THE TWO MORPHOLOGICAL MODELS}

Both the GJ14 and RL08 models were constructed to fit the \citet{Redfield2002}
data set of \ion{Mg}{2} and \ion{Fe}{2} absorption line radial velocities, although
the RL08 model was constructed to also fit radial velocities for other 
lines of sight measured from \ion{Ca}{2} line spectra. A critical test of 
the viability of both models is, therefore, to determine how
accurately each model predicts the radial velocities of a new
data set with
targets randomly distributed in Galactic coordinates. 

M14 obtained high-resolution STIS 
E230H spectra of the \ion{Mg}{2},
\ion{Fe}{2}, and \ion{Mn}{2} lines in the 2580--2805~\AA\ spectral region 
for stars that previously had only 
1200--1700~\AA\ spectra. This new data set
consists of 76 velocity components measured in the lines of sight
toward 34 stars.  To be consistent with the RL02 sample, we limit our
analysis to stars within 100 pc, and therefore only used 32 of the 34
sight lines from
this sample.  The data were obtained in the SNAPshot observing
mode---designed to provide short (typically one spacecraft orbit)
observations to fill gaps in the HST schedule. Since the observed
targets were selected by the HST schedulers from a large list of stars
distributed across the sky, the observed targets are, in effect, randomly
distributed in Galactic coordinates as shown in Figure 1 in M14. 
The M14 data set, therefore, provides us with an appropriate basis for
testing the two models. 

We will evaluate how well the GJ14 and the RL08
models fit the new data and utilize some statistical tests to make 
quantitative comparisons,
however statistical tests have their limitations as the fitting
parameters for the two models are very different and could lead to possible
systematic errors.
Figure 1 compares the observed minus predicted radial velocities for the two
models as a function of Galactic
longitude and latitude using the M14 data set. 
The left panels separate by color 
the predicted velocities for the Local Cloud (32 measurements) and Cetus
Ripple (9 measurements) using the
procedure proposed by
GJ14. The right panels make a similar comparison for the exact
  same set of velocity components, except using instead the RL08
model for their suite of discrete clouds.  Note that two components associated
  with the Cetus Ripple by GJ14 are not identified with a known cloud
  in the RL08 model resulting in $32+9 = 41$ total points in the GJ14
  comparison and $32+7=39$ points in the RL08 comparison.  For
    the latter comparison, we use
  a weighted mean velocity based on all available LISM
  measurements, which for the M14 sample is primarily \ion{Mg}{2} 
  and \ion{Fe}{2}.  This mitigates systematic errors that can arise 
  from using a single ion (e.g., due to saturation or wavelength 
  calibration issues).

Figure 2 shows a similar set of comparisons, but now using the
RL08 dataset. This dataset includes observations tabulated in 
\citet{Redfield2002} and \citet{Redfield2004}, totaling $\sim$80 sight 
lines.  The GJ14 interpretation of the data results in 80 Local Cloud 
and 36 Cetus Ripple assignments, totally 116 components.  The RL08 
interpretation of the data results in 122 cloud identifications.      
This comparison using this full sample is not as clear of a test of the two
morphological models because both models were constructed to fit these
data (together with some \ion{Ca}{2} data for the RL08 model). Nevertheless, this
second comparison is
useful as a confirmation of the results of the first comparison. 

As shown in Figures 1 and 2, the RL08 model clearly provides a much
tighter fit to both the M14 and RL08 data sets.  
The root mean square (RMS) scatter for the RL08 model is a factor 
of 1.5 or two times smaller than for the GJ14 model: 1.4 versus
  2.1  for the M14 dataset and 0.77 versus 1.5 for the RL08 data set.  
However, the RL08 model has 45 free parameters (3 for each of the 15 clouds), 
versus the 8 free parameters in the GJ14 model. For the GJ14 model, there are 
3 parameters for the Local Cloud velocity vector, an
additional 3 for the deformation (direction and magnitude), and 2 
for the Cetus Ripple, which is just
an offset and range around that offset. An $F$-test shows that
even with more free parameters, the RL08 model is statistically
preferred.  If we look at the entire data set, that is data included 
in RL08 and M14, the GJ14 model has a reduced $\chi^2_\nu$ of 49.5, with 
149 degrees of freedom ($157-8$), whereas the {\it same} 157
components in the RL08 model has a
reduced $\chi^2_\nu$ of 30.4, with 112 degrees of freedom ($157-45$).  
There is a 0.4\% probability that such a dramatic difference in 
the ratio of $\chi^2$ could be the result of a random set of data as 
compared to the true model, even taking into account the higher number 
of parameters for the RL08 model \citep{Bevington1992}.  
Note that the $\chi^2$ calculations above use the individual,
published 
velocity errors, whereas GJ14 implemented a velocity error floor 
of 1.1 km~s$^{-1}$.  If we make the same assumption regarding the 
velocity errors, the reduced $\chi^2_\nu$ values become 2.30 and 0.92 
for the GJ14 model and RL08 model, respectively.  However, the 
basic conclusion remains the same, that the difference in the 
quality of fit between the two models is significant.  

Of the 210 velocity
components (147 from the RL08 data sample and 63 from the
M14 sample, which excludes those $>$100 pc and a component identified 
with absorption from a circumstellar disk), the RL08 model is able 
to successfully fit 175 (83\%), whereas the GJ14 model fits 157
(75\%).  While the GJ14 model must have a Local Cloud component along 
every sight line, in practice, essentially all sight lines have at 
least one successfully assigned component in the RL08 model.  On the 
other hand, the GJ14 model is unable to account for sight lines with 
more than two absorption components (Local Cloud and Cetus
Ripple), whereas the RL08 model can
account for them with several clouds along the line of sight.
Although the velocity component assignment percentages are similar 
between the two models, the predictive power of the RL08 model is much
better than the GJ14 model when tested against the M14 data set 
and the entire data set as a whole.

For all 34 stars observed and analyzed by M14, the RL08 kinematic model 
predicts 40 absorbers along the lines of sight.  All 40 radial
velocity predictions are successfully detected in the M14 sample to 
within measurement uncertainty (i.e., within $\sim$3$\sigma$). The 
RL08 kinematic model also accurately predicts the
observed radial velocities for 18 absorbers along sight lines lying 
within $20^{\circ}$ of a cloud boundary.  The RL08 model clouds have 
the condition of being contiguous, and have a spatial extent (as does 
the Cetus Ripple in the GJ14 model).  Given the relative sparseness 
of the sample, the precise cloud morphologies are not necessarily 
well constrained, but will be refined as new data is made available 
as long as the cloud boundary remains contiguous.  In their 
high-resolution STIS spectra of three early B stars located
about 70 pc from the Sun, \citet{Welsh2010a} also found that 8 out of
11 velocity components for these lines of sight are consistent with
the predicted velocities of six clouds in the RL08 kinematic 
model. The 40 out of 40 success rate in ascribing M14 radial
velocities to clouds confirms the
robustness of the RL08 model, although any model consisting of
rigid flow structures is an approximation to what is likely a more
complex flow pattern that could include rotation in addition to
nonrigid translation terms. 
It is likely that a more realistic description of the 
CLIC kinematics lies somewhere between the two extreme morphologies.

\begin{figure}
\centerline{\includegraphics[width=6.5in, angle=180, scale=1.0]
{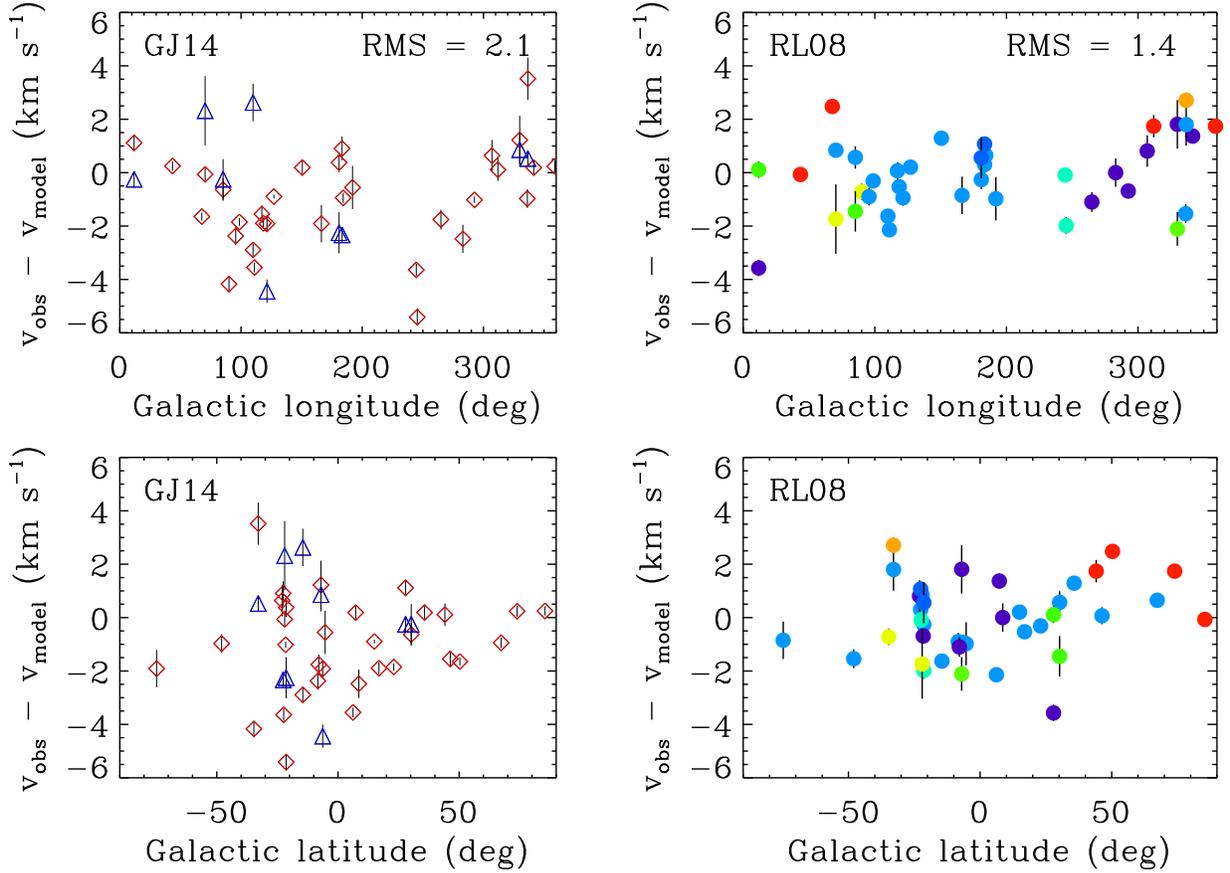}}
\caption{
Comparison of the velocity offsets (observed minus model prediction)
for the M14 data set. Left panels use the GJ14 model
(red for Component 1 and blue for the Cetus Ripple component).
Right panels use the RL08 model (the color coding is by cloud
structure, which is dominated by 
the LIC [blue], G Cloud [violet], and NGP Cloud [red]).  
The root mean scatter (RMS) is given in the top right corner, 
and is a factor of 1.5 lower for the RL08 model than 
the G14 model.}
\end{figure}

\begin{figure}
\centerline{\includegraphics[width=6.5in, angle=180, scale=1.0]
{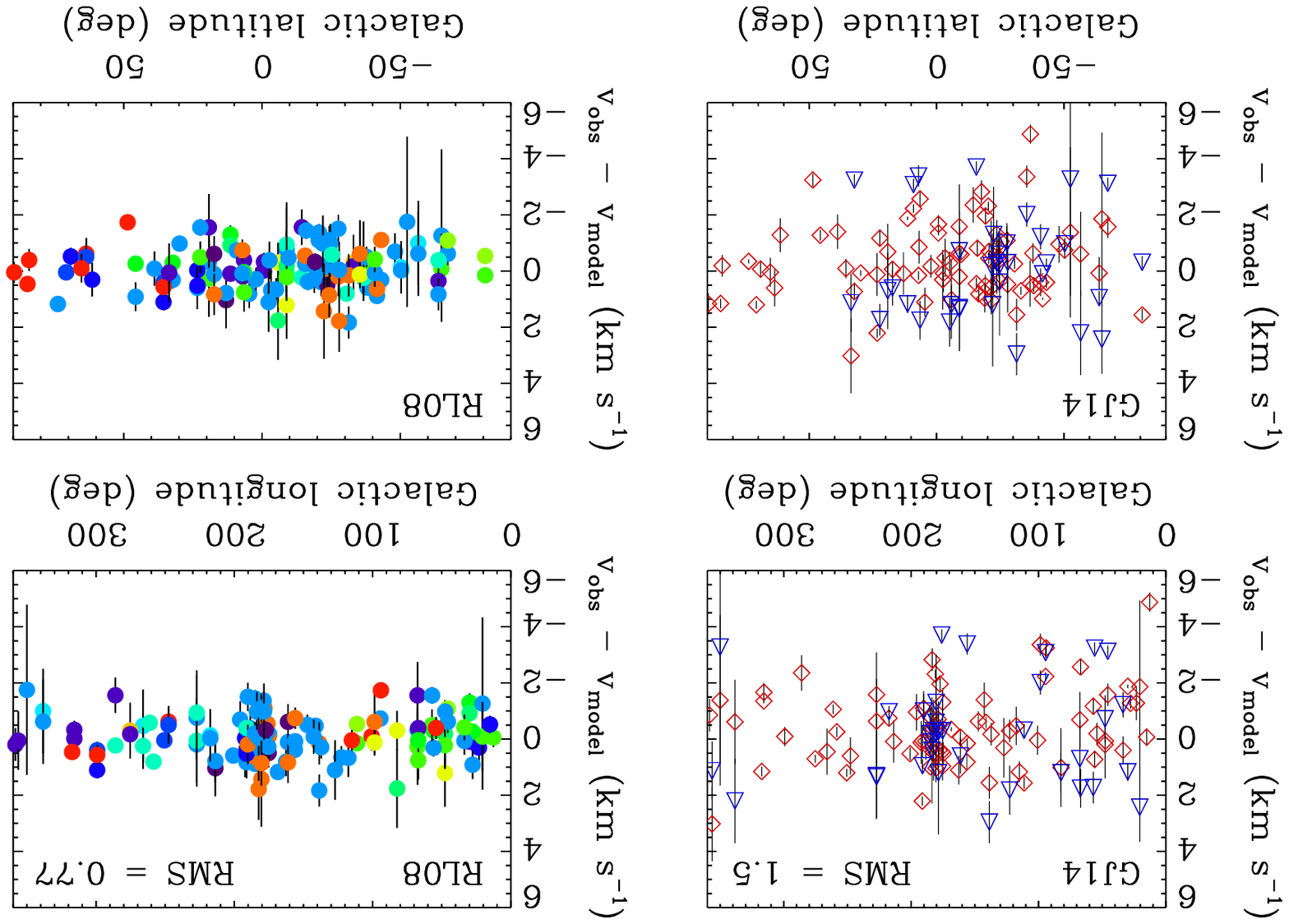}}
\caption{
Comparison of the velocity offsets (observed minus model prediction)
for the RL08 data set. Left panels use the GJ14 model
(red for Component 1 and blue for the Cetus Ripple
component). Right panels use the RL08 model (the color coding is by 
cloud structure, which is dominated by 
the LIC [blue], Hyades Cloud [orange], G Cloud [violet], Mic Cloud
[green], Blue Cloud [dark blue], and NGP Cloud [red]).  
The RMS scatter is given in the top right corner, and is a
factor of two lower for the RL08 model than the G14 model.}
\end{figure}

\section{DISCUSSION: WHICH MORPHOLOGICAL MODEL IS MORE REALISTIC AND WHY IS 
THIS IMPORTANT?}

Whether the warm partially ionized interstellar gas located within a
few parsecs of the Sun has a continuous or a discrete multicloud
morphology is important because the morphology can provide important 
clues concerning 
the physical properties of the gas and its evolution. 

The presence of discrete clouds requires a 
structuring agent that could be ionizing radiation from an assorted 
distribution of different hot stars
shielded by clouds in some directions, inhomogeneous magnetic fields,
shock waves, or thermal instabilities. If the warm gas is indeed structured
into discrete separated clouds, then there must be gas located between
the clouds that is difficult to detect in spectral lines of
hydrogen and singly-ionized metals. 
This intercloud gas has not been unambiguously characterized perhaps because it is
very hot (roughly $10^6$~K) as originally suggested by the observed 
diffuse soft X-ray emission \citep[cf.][]{Cox2005}, and recently confirmed by \citet{Galeazzi2014}.  However, a significant contribution to this soft X-ray emission results from charge-exchange emission inside of the heliosphere 
\citep{Snowden1994, Koutroumpa2012}. Components of the intercloud
medium could 
be comprised of highly ionized recombining gas \citep{Welsh2009} or 
Stromgren sphere $10^4$~K ionized hydrogen gas surrounding nearby hot
stars like Sirius~B \citep{Tat1999}. An advantage of the continuous 
cloud morphology model is that it does not require interstitial gas and is therefore 
consistent with its nondetection, whereas discrete detached multicloud models require the
presence of such gas and, if/when detected, the properties of this gas
will provide insight into the energy balance of the CLIC.

Whether the warm cloud gas fills a small fraction
or essentially all of the volume within about 9~pc of the Sun
depends upon the neutral hydrogen gas density. If n(HI)$\approx
0.20$ cm$^{-3}$ is typical for the warm gas in nearby clouds, 
as \citet{Slavin2008}
find for the LIC, then the warm partially ionized clouds fill a small 
fraction of nearby space. If, on the other hand, 
n(HI)$\approx 0.055$ cm$^{-3}$, as suggested by GJ14,
then the warm gas fills all of nearby space and there is no
interstitial gas within about 9 parsecs of the Sun.

\citet{Gry2014} note that if the LISM is comprised of discrete clouds
and the filling factor is low, it is likely that sight lines could be
found that do not traverse any warm gas and, therefore,
show no LISM absorption.  However, every sight line analyzed so
far shows detected interstellar absorption with $\log N($\ion{H}{1}$)$ $>17.4$.
\citet{Redfield2008} reported a filling factor of 5.5--19\%  
for their suite of 15 clouds, but this calculation 
assumed the maximum possible spacing of the clouds (limited to the
closest stars that showed absorption for a particular cloud) and the 
placement of the clouds within 15 pc of the Sun. Also, 
\cite{Welsh2010a} argued that the clouds are all located within 10~pc as
there is little warm gas between 10~pc and the 70~pc location of
their targets. In order to explore the question of why there are
not gaps in \ion{H}{1} absorption, we recomputed the suite of simulations 
described in \citet{Redfield2008} while keeping track of the 
probability of any given sight line avoiding all simulated clouds.
Our objective was to find an upper limit to the total volume enclosing
the LISM clouds that predicts that essentially all lines of sight to
nearby stars have observable HI absorption. We made several 
simplifying assumptions in this analysis and our conclusions are based 
on a statistical sample of simulated LISM configurations.  A more accurate 
estimate would require a full 
three-dimensional model of the LISM based on the entire dataset.  
We find that if all the clouds are located within 7 pc of the
Sun, the filling factor ranges from 55--100\% and there is 
a $\sim$99.6\% probability that all sight lines will traverse at 
least one absorbing cloud (using the original 15 pc limit, this 
probability is $\sim$65\%).  We conclude that if the 15 clouds 
in the RL08 model are located within 7 pc, there should be few or 
no gaps in the projection of these clouds on the sky and no sight 
lines to nearby stars with undetected \ion{H}{1} absorption. 
Note that this conclusion is strongly dependent on the number 
of very nearby stars (e.g,. $<$5 pc), and an observational campaign 
to expand this sample could be a powerful discriminant between these two models.
While both the GJ14 and RL08 models
largely fill the immediate interstellar space 
surrounding the Sun, the kinematic properties of the two
models are quite distinct, 
which may hold clues regarding the origin and evolution of this material.

There is evidence that the interstellar magnetic field strength 
immediately surrounding the heliosphere lies
in the range 2.7--5 $\mu$G \citep{Frisch2011}. 
Since the beginning of 2013, the Voyager 1 spacecraft has been 
making {\em in situ} measurements of the interstellar 
magnetic field strength outside of the termination shock 
with a mean value of $4.64\pm0.09 \mu$~G 
and nearly constant orientation very different from that of 
the solar magnetic field inside of the heliosheath 
\citep{Gurnett2013,Burlaga2014a,Burlaga2014b}. 
Although the measured magnetic field likely
refers to the field draped around the heliopause, which may be
compressed relative to the 
undisturbed field far away from the Sun, Voyager 1 has provided us with 
the best available estimate of the interstellar magnetic field
strength near the Sun. Since equipartition between magnetic and thermal
energy is about 2.7 $\mu$G in the LIC, the Voyager 1 magnetic field
strength measurement indicates that strong interstellar magnetic 
fields are present near the Sun and could, therefore, control the 
structure of discrete clouds. The thin elongated structures of several
clouds identified by RL08 (Aur, Cet, Mic, and perhaps others) are
consistent with confinement by elongated parsec scale magnetic fields. 

Large-amplitude intraday and annual scintillations of some 
well monitored, unresolved quasars at
radio wavelengths indicate turbulent-scattering screens in the
CLIC. \cite{Linsky2008} showed that the variability of three
well-studied quasars (B1257-326, B1519-273, and J1819+385) can be
understood as produced by the Earth's orbital motion through the diffraction
pattern of scattering screens located within 7 parsecs of the Sun. The 
lines of sight to these quasars pass close to the outer edges 
of two or more
adjacent or perhaps colliding clouds. The shear of the different
cloud velocities and the likely higher ionization of the gas at the 
cloud edges due to external ionizing radiation or thermal conduction from
hot surrounding gas could produce the turbulent ionized plasma 
and the inhomogeneous index of refraction properties of
scattering screens.  The
transverse velocities measured for these scintillation screens match
very well with a cloud in their line of sight for
the RL08 model but poorly for gas at the
Local Standard of Rest.  The nonridgid structure of the GJ14 model
is consistent with a scattering screen in
the J1819+385 sight line, but cannot explain the scattering screens
for the other two sight lines. The existence of nearby scattering screens 
provides additional evidence that isolated warm
clouds with ionized edges rather than a single cloud
even with a nonridgid velocity structure can more naturally
explain the complexity observed in the LISM velocity structure.

Another test of whether or not warm partially-ionized gas is confined 
in identifiable clouds, would be the observation of
absorption by high ionization species located near
the outer boundaries of clouds produced by either thermal conduction from
surrounding hot gas or ionization by the extreme ultraviolet
radiation from $\epsilon$~CMa and other sources. Searches for
\ion{O}{6} $\lambda 1032$ absorption \citep[e.g.,][]{Savage2006}  
led to detections in hot white
dwarf photospheres but not in the interstellar medium. 
In their summary of searches for lower stages of ionization in the
LISM, \citet{Welsh2010b} found that the only clear example is 
interstellar \ion{C}{4} absorption in the line of sight to
the B2Ve star HD~158427 located at a distance of $~\sim74$~pc 
inside of the Local Cavity. This detection is very interesting because
the observed radial velocity, $-24.3\pm 2.0$ km~s$^{-1}$, is consistent
with the projected radial velocities of both the G and Aquila clouds
and the line of sight to HD~158427 is tangential to the edges of both 
clouds as shown in RL08. This is the most favorable geometry for the
detection of weak absorption, because of the long line of sight through
the cloud edges.

\section{CONCLUSIONS}

The M14 data base of interstellar radial velocities provides
an unbiased test of the robustness of two fundamentally different
morphological models of the interstellar medium within a few parsecs
of the Sun. Both the multiple discrete cloud model proposed by 
RL08, for which each cloud has a rigid flow vector,
and the single local cloud model with nonrigid flows
proposed by GJ14 are approximations to what is likely a more complex 
kinematic and morphological structure. Nevertheless, it is
instructive and important to
test the predictive power of both models. We find that the RL08
model fits the new velocity data significantly better than the GJ14
model and provides a natural way of explaining the observed
multiple velocity components along the lines of sight to
many nearby stars. 
Also, the likely presence of strong magnetic fields and scattering
screens in the CLIC are arguments for the presence of
multiple clouds that do not entirely fill nearby space, although a 
close packing of clouds is likely, resulting in a high filling factor 
and a low probability of a sight line displaying no observable 
LISM absorption. 
Important future tests of the CLIC morphology would be
the detection and identification of interstitial gas, if it is
present, and the measurement of neutral hydrogen densities far 
away from the heliosphere.  A more densely sampled observational data 
set, particularly comprised of very close stars (e.g., $<$5 pc), would 
also provide definitive tests between smoothly varying
deformed kinematics of a continuous medium and the more striking 
variations that would arise in a suite of discrete absorbers.  
We encourage future model development and
testing to better understand the physical properties and evolution of
the interstellar medium close to the solar neighborhood.

We acknowledge support through NASA HST grant GO-11568 from the Space
Telescope Science Institute, which is operated by the Association of
Universities for Research in Astronomy, Inc. for NASA under contract 
NAS5-26555. We thank Drs. Cecile Gry and Edward Jenkins for their careful
reading and in depth comments concerning an earlier version of this
paper that guided us in completing this paper.  We appreciate the expeditious and insightful review by the referee.

{\it Facilities:} \facility{HST (GHRS, STIS)}

\end{document}